\documentclass[a4paper]{article}
\usepackage{INTERSPEECH2022}
\usepackage[utf8]{inputenc}
\usepackage{url}
\usepackage{booktabs}
\usepackage{array}
\usepackage{multirow}
\usepackage{xcolor}
\usepackage{tikz}
\usetikzlibrary{shapes.geometric,arrows.meta,positioning,calc}
\tikzset{
  pbox/.style={draw,rounded corners=2pt,align=center,inner sep=3pt,
    font=\scriptsize,text width=68mm,minimum height=6mm,fill=black!3},
  pio/.style={draw,align=center,inner sep=3pt,font=\scriptsize\bfseries,
    text width=40mm,minimum height=5mm,fill=black!8},
  pfind/.style={draw,rounded corners=2pt,align=left,inner sep=4pt,
    font=\scriptsize,text width=68mm,fill=blue!4,draw=blue!40},
  parr/.style={-{Latex[length=2mm]},thick}
}

\title{An Omnilingual-ASR-Based Speech-LLM System for the 2nd MLC-SLM Challenge}

\name{Shuming Fang$^1$\thanks{Accepted to INTERSPEECH 2026.}, Shuifei Zeng$^2$}
\address{$^1$UGREEN AI Lab, Shenzhen, China\\
$^2$School of Computer Science (National Pilot Software Engineering School),\\
Beijing University of Posts and Telecommunications, Beijing, China}
\email{simon.fang@ugreen.com}

\begin{document}

\maketitle

\begin{abstract}
We describe our submission to Task~1 of the 2nd MLC-SLM Challenge:
a cascaded diarization-then-recognition system that combines
DiariZen-Large-s80 (WavLM-Large) segmentation, CAM++ embedding-based
two-speaker clustering, and a LoRA-adapted \texttt{omniASR\_LLM\_7B\_v2} recognizer, with no oracle segmentation
or speaker labels at test time. On the official Development set
(150 conversations, 21 language/accent categories) the system attains a
macro tcpMER of \textbf{29.27\%}, versus \(79.15\%\) for the official
baseline; on the Evaluation set it scores \(50.23\%\).
We also analyze two engineering choices that substantially affect
tcpMER.
First, embedding-based speaker clustering outperforms an
end-to-end-style alternative that assigns speakers from ASR
\texttt{<sc>} turn markers alone.
Second, overlap-aware segmentation,
although intended to raise diarization recall, \emph{increases} tcpMER
because overlapped speech is transcribed twice.
\end{abstract}

\section{Introduction}
Conversational speech in realistic, multilingual settings remains
challenging for both automatic speech recognition (ASR) and speaker
diarization (SD). The Multilingual Conversational Speech Language Model
(MLC-SLM) Challenge~\cite{mlcslm2025} targets this gap by releasing a
real-world, multi-language conversational corpus. In its second edition,
\textbf{Task~1} asks participants to jointly perform speaker diarization
(``who spoke when'') and recognition (``what was said'') on raw
recordings, \emph{without} any oracle segmentation or speaker labels at
evaluation time. Systems are \textbf{ranked} by the time-constrained
minimum-permutation error rate (tcpMER), i.e.\ tcpWER for most languages
and tcpCER for Japanese, Korean and Thai~\cite{vonneumann2023meeteval};
speaker permutations are resolved before concatenating transcripts for
scoring. The official Task-1 baseline
fine-tunes Microsoft's open-source VibeVoice-ASR with LoRA on the
challenge training set~\cite{mlcslm2_task1_baseline}.

This paper describes our Task~1 submission and the Development-set
ablations used to select it. The system is a cascaded pipeline---neural
segmentation, CAM++-based speaker clustering, and LoRA-adapted
ASR---so that each stage can be tuned and compared under the official
tcpMER protocol. The remainder of the paper is organized as follows:
Section~2 details the implementation, Section~3 the evaluation setup,
and Section~4 reports Development results, component ablations
(Table~\ref{tab:ablation}), and the Evaluation-set leaderboard score.
In summary:
\begin{itemize}\itemsep2pt
  \item We \textbf{describe} the final cascaded system submitted to the
  challenge (Exp4 in Table~\ref{tab:ablation}).
  \item We \textbf{compare} embedding-based clustering with end-to-end
  \texttt{<sc>}-based speaker assignment on the Development set.
  \item We \textbf{compare} overlap-enabled and overlap-disabled
  segmentation fronts under the same scoring protocol.
  \item We \textbf{report} the Evaluation-set leaderboard score and
  discuss the gap to Development tuning.
\end{itemize}

\section{System Description}
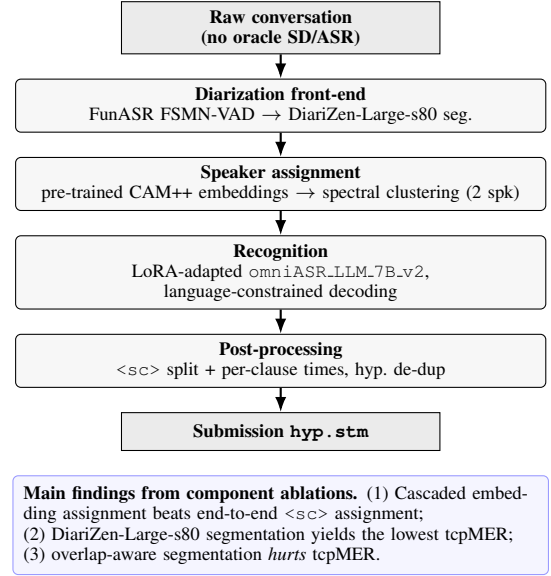
\begin{figure}[t]
  \centering
  \begin{tikzpicture}[node distance=3.2mm]
    \node[pio] (in) {Raw conversation (no oracle SD/ASR)};
    \node[pbox,below=of in] (seg)
      {\textbf{Diarization front-end}\\ FunASR FSMN-VAD
       $\rightarrow$ DiariZen-Large-s80 seg.};
    \node[pbox,below=of seg] (emb)
      {\textbf{Speaker assignment}\\ pre-trained CAM++ embeddings
       $\rightarrow$ spectral clustering (2 spk)};
    \node[pbox,below=of emb] (asr)
      {\textbf{Recognition}\\ LoRA-adapted \texttt{omniASR\_LLM\_7B\_v2},
       language-constrained decoding};
    \node[pbox,below=of asr] (post)
      {\textbf{Post-processing}\\ \texttt{<sc>} split + per-clause times,
       hyp.\ de-dup};
    \node[pio,below=of post] (out) {Submission \texttt{hyp.stm}};
    \node[pfind,below=of out] (find)
      {\textbf{Main findings from component ablations.}
       (1)~Cascaded embedding assignment beats end-to-end
       \texttt{<sc>} assignment;\\
       (2)~DiariZen-Large-s80 segmentation yields the lowest tcpMER;\\
       (3)~overlap-aware segmentation \emph{hurts} tcpMER.};
    \draw[parr] (in) -- (seg);
    \draw[parr] (seg) -- (emb);
    \draw[parr] (emb) -- (asr);
    \draw[parr] (asr) -- (post);
    \draw[parr] (post) -- (out);
  \end{tikzpicture}
  \caption{Overview of the submitted cascaded pipeline (top to bottom) and
  the main outcomes of our Development-set ablations (Table~\ref{tab:ablation}).}
  \label{fig:pipeline}
\end{figure}

Figure~\ref{fig:pipeline} summarizes the pipeline. A raw conversation is
(1)~segmented into single-speaker regions, (2)~assigned speaker identities
by embedding extraction and clustering, (3)~transcribed
segment-by-segment by the ASR model, and (4)~cleaned and packed into the
submission format. The four stages are decoupled so that each can be
replaced and ablated independently.

\subsection{Speaker diarization front-end}
\label{ssec:diar}
We build on the 3D-Speaker toolkit~\cite{3dspeaker} with a cascaded
layout: \emph{segmentation} is handled separately from \emph{speaker assignment}.
This separation matches our
implementation and makes the Exp1--Exp4 comparisons in
Section~\ref{sec:results} straightforward---each stage can be swapped
without retraining the recognizer.

\noindent\textbf{Voice activity detection.} All diarization configurations
first apply the FunASR feedforward sequential memory network (FSMN)
voice-activity detector~\cite{gao2023funasr,funasr_fsmn_vad}, using the
ModelScope checkpoint
\texttt{speech\_fsmn\_vad\_zh-cn-16k-common-pytorch}. VAD is a shared
module in every inference run and is not part of the segmentation-model
comparison below.

\noindent\textbf{Segmentation.} On top of VAD we compare two publicly
released neural segmenters, both integrated via the 3D-Speaker
\texttt{pyannote.audio} inference stack~\cite{3dspeaker}: (a)~the
HuggingFace checkpoint \texttt{pyannote/segmentation-3.0}~\cite{plaquet2023powerset,bredin2023pyannote},
a \textbf{PyanNet} powerset segmentation model (SincNet + BiLSTM head;
\(10\,\)s windows, \(10\%\) hop) used as a segmenter only---we do
\textbf{not} use the pyannote speaker-embedding pipeline, and instead
pair it with CAM++ clustering below; overlapped-speech decoding from the
powerset head is enabled in Exp2 and disabled in Exp3; and (b)~the
\textbf{DiariZen-Large-s80} checkpoint (\texttt{BUT-FIT/diarizen-wavlm-large-s80-md}),
a WavLM-Large + Conformer EEND segmentation model~\cite{han2025diarizen},
run with \(12\,\)s windows and a \(10\%\) hop (Exp4, submitted). The raw
segments are post-processed by (i) dropping segments shorter than
\(0.7\,\)s, (ii) merging same-speaker segments separated by gaps
\(\le 1.5\,\)s, and (iii) splitting any segment longer than
\(25\,\)s, which keeps the per-segment duration within the ASR model's
stable input range.

\noindent\textbf{Speaker embedding.} We use the publicly released,
pre-trained \textbf{CAM++}~\cite{wang2023campp} checkpoint from the
3D-Speaker toolkit~\cite{3dspeaker}
(\texttt{campplus\_cn\_en\_common}): it is lightweight, supports
multilingual telephone speech out of the box, and fits our two-speaker
clustering stage without additional speaker-model training. For each
candidate segment we slice \(1.5\,\)s sub-windows with a \(0.75\,\)s hop
and extract \(192\)-dimensional embeddings from an \(80\)-dimensional
log-filter-bank front-end.

\noindent\textbf{Clustering.} All sub-window embeddings of a recording are
grouped by spectral clustering~\cite{park2020spectral} with the number of
speakers fixed to two, matching the two-party telephone conversations of
the corpus. Each candidate segment then receives a single anonymous
speaker ID (\texttt{Speaker1}/\texttt{Speaker2}), and adjacent
same-speaker segments are merged. Because the DiariZen
front-end can hypothesize concurrent speakers, a further post-processing
step resolves any residual cross-speaker time overlap (the shorter segment
yields the overlapped region), so the segments passed to the recognizer
are effectively non-overlapping.

\subsection{Multilingual ASR with LoRA adaptation}
\label{ssec:asr}
The recognizer is Meta's publicly released
\texttt{omniASR\_LLM\_7B\_v2} checkpoint from the Omnilingual
ASR family~\cite{omnilingual2025} (a wav2vec2-style encoder with a CTC
head). We adapt it to the conversational, telephone-channel domain with
Low-Rank Adaptation~\cite{hu2022lora} rather than full fine-tuning: with a
single GPU and a 7B backbone, LoRA is the only practical option, and
freezing the acoustic encoder also guards a strong multilingual model
against over-fitting the comparatively small in-domain set. Only the
low-rank adapters (rank \(r{=}8\), \(\alpha{=}16\), dropout \(0.05\)) are
updated, under a CTC objective. Adaptation runs in fairseq2 for \(50\)k
steps with FSDP and bfloat16; utterances are filtered to \(2\)--\(15\,\)s
and length-bucketed (up to \(4{\times}10^{5}\) audio elements per batch,
gradient accumulation \(16\)). We use AdamW (lr \(2{\times}10^{-4}\),
\(\beta{=}(0.9,0.98)\), no weight decay) with a tri-stage warm-up/hold/decay
schedule (ratios \(0.1/0.4/0.5\)). All reported systems use the same
\(50\)k-step checkpoint.
Training utterances are packed from consecutive reference-segmented
turns in the challenge training set (oracle boundaries available only
during adaptation, not at test time) subject
to the length filter; transcripts concatenate segment text and insert a
speaker-change token (\texttt{<sc>}) between adjacent segments from
different speakers (but not between consecutive segments of the same
speaker), so the model learns to predict turn boundaries alongside
lexical content.

At inference, each diarized segment is decoded with the target language
supplied explicitly as a decoding constraint
(\texttt{lang=<lang>}), which confines the decoder to the target
language's token space and effectively eliminates spurious
language switching.

\subsection{Post-processing}
\label{ssec:post}
Two inference-time steps reduce time-constrained errors without changing
the acoustic model. (1)~\emph{Speaker-change splitting}: the recognizer
outputs the \texttt{<sc>} markers it was trained on; within a segment we
split on these markers and allocate per-sub-clause timestamps proportional
to character length, instead of sharing a single timestamp across the
whole segment. (2)~\emph{Repeated-hypothesis de-duplication}: within a
segment, near-duplicate sub-clauses (identical, substring, or
\(\ge\!0.9\) sequence similarity after NFKC + case folding) are removed to
suppress decoder echo. All Development-set numbers in this paper,
including Exp1--Exp4, are scored with the \emph{same} official symmetric
normalization applied to references and hypotheses (NFKC, then strict
tokenization aligned with the challenge WER/CER scripts, with punctuation
removal for word cohorts); see Section~\ref{sec:setup}.

\section{Experimental Setup}
\label{sec:setup}
\noindent\textbf{Data.} All results are reported on the official MLC-SLM
\emph{Development} set: 150 long conversational recordings
(\(\sim\!30\) min telephone-channel sessions) spanning 21
language/accent categories (English is split into five accents, following
the official breakdown). The final system is submitted on the
\emph{Evaluation} set. LoRA adaptation uses the official challenge
training manifest with reference turn boundaries (available for training
only).

\noindent\textbf{Metrics.} Following the challenge, we report tcpWER for
18 categories and tcpCER for Japanese, Korean and Thai, with a
collar of \(5\,\)s computed by MeetEval~\cite{vonneumann2023meeteval}.
For each cohort we report the \emph{micro} rate (total errors over total
reference tokens); the overall \emph{macro} score is the mean of
per-category micro rates over all 21 categories. \textbf{All results
reported in this paper}---main results, Exp1--Exp4, and baseline
comparisons---apply this \emph{identical} scoring-time normalization
(Unicode NFKC compatibility normalization, which maps full-width and other
compatibility-variant characters to their canonical forms, followed by
strict tokenization aligned with the challenge WER/CER scripts, with
punctuation removal for word cohorts) symmetrically to references and
hypotheses, so neither side gains from one-sided text cleaning. The
official baseline Development scores (Table~\ref{tab:baseline}) are quoted
from the released repository~\cite{mlcslm2_task1_baseline}, where the
authors likewise apply their \texttt{text\_normalization\_2nd.py} (NFC,
lowercasing, punctuation removal) symmetrically to both sides under the
same collar and tcpMER protocol; we do not re-score their system. The
\(\sim\!50\) point gap to our \(29.27\%\) therefore reflects
architecture, not asymmetric normalization.

\noindent\textbf{Hardware.} ASR LoRA adaptation was performed on a single
NVIDIA RTX~6000 Ada GPU.

\section{Results and Analysis}
\label{sec:results}

\subsection{Main results}
Table~\ref{tab:perlang} reports per-language results of the final system
(DiariZen-Large-s80 segmentation + pre-trained CAM++ + two-speaker clustering +
LoRA-adapted \texttt{omniASR\_LLM\_7B\_v2}). Following the official scoring protocol we
report over all \(21\) language/accent categories (English is broken down
by accent). The system attains a macro tcpMER of
\textbf{29.27\%}, with a word-cohort micro tcpWER of \textbf{29.41\%} and
a character-cohort micro tcpCER of \textbf{27.38\%}. For reference, the
official Task-1 baseline---a LoRA-fine-tuned VibeVoice-ASR
recognizer~\cite{mlcslm2_task1_baseline}---reports a macro tcpMER of
\(79.15\%\) on the same Development set under the identical 21-category
protocol; our cascaded system lowers this to \(29.27\%\), a relative
reduction of roughly \(63\%\).

\begin{table}[t]
\centering
\caption{Per-language results of the final system on the Development set,
following the official \(21\)-category breakdown (English reported by
accent). Japanese/Korean/Thai are scored with tcpCER, the rest with
tcpWER (micro, \%).}
\label{tab:perlang}
\small
\begin{tabular}{@{}ll@{\hspace{1.2em}}ll@{}}
\toprule
Language & tcpMER & Language & tcpMER \\
\midrule
Spanish (MX)        & 11.1 & Eng. (British)     & 27.8 \\
Eng. (Australian)   & 17.8 & Vietnamese         & 33.0 \\
Eng. (Indian)       & 18.2 & French             & 34.6 \\
Thai$^\dagger$      & 18.7 & Korean$^\dagger$   & 34.8 \\
Italian             & 19.9 & Tagalog            & 34.9 \\
Spanish             & 20.0 & German             & 37.8 \\
Eng. (Filipino)     & 20.1 & Portuguese         & 39.4 \\
Portuguese (BR)     & 20.3 & Japanese$^\dagger$ & 40.3 \\
Russian             & 22.9 & French (CA)        & 46.7 \\
Urdu                & 24.0 & Turkish            & 65.8 \\
Eng. (American)     & 26.6 &                    &      \\
\midrule
\multicolumn{4}{@{}l}{Word cohort micro tcpWER (18 categories): \textbf{29.41}} \\
\multicolumn{4}{@{}l}{Char cohort micro tcpCER (3 categories): \textbf{27.38}} \\
\multicolumn{4}{@{}l}{Macro tcpMER (21 categories): \textbf{29.27}} \\
\bottomrule
\end{tabular}
\par\vspace{2pt}
{\footnotesize $^\dagger$ scored with tcpCER. The five English accents
average \(22.2\%\).}
\end{table}

\subsection{Comparison with the official baseline}
For context, Table~\ref{tab:baseline} lists the per-language Development-set
scores of the official Task-1 baseline as released in the repository of
\cite{mlcslm2_task1_baseline}; the baseline is a
LoRA-fine-tuned VibeVoice-ASR recognizer evaluated under the same
\(5\,\)s collar with tcpCER for Japanese/Korean/Thai and tcpWER
elsewhere, with symmetric text normalization as above. Its scores are high
across the board---no single category falls
below \(60\%\) and the easiest cohorts still sit above \(63\%\)---yielding
a \(79.15\%\) average over its \(21\) language/accent categories. Using the
\emph{same} 21-category protocol, our cascaded system reaches a macro
tcpMER of \(29.27\%\); even our hardest cohort (Turkish, \(65.8\%\)) is
comparable to the baseline's \emph{single best} category (\(63.4\%\)).
This gap indicates that the combination of a strong segmentation
front-end, embedding-based two-speaker clustering, and a LoRA-adapted
multilingual recognizer is markedly more effective than a single
fine-tuned speech-LLM recognizer on this conversational, telephone-channel
data.

\begin{table}[t]
\centering
\caption{Official 2nd MLC-SLM Task-1 baseline (LoRA-fine-tuned
VibeVoice-ASR) per-language tcpMER on the Development set, as reported in
the official baseline repository~\cite{mlcslm2_task1_baseline} (collar \(5\,\)s; tcpCER for
Japanese/Korean/Thai, tcpWER otherwise). For comparison, our system attains
a macro tcpMER of \textbf{29.27\%} under the same 21-category protocol.}
\label{tab:baseline}
\small
\begin{tabular}{@{}ll@{\hspace{1.2em}}ll@{}}
\toprule
Language & tcpMER & Language & tcpMER \\
\midrule
Eng. (American)    & 77.39 & Portuguese        & 75.64 \\
Eng. (Australian)  & 81.50 & Portuguese (BR)   & 73.02 \\
Eng. (British)     & 67.60 & Russian           & 83.84 \\
Eng. (Filipino)    & 63.36 & Spanish           & 82.51 \\
Eng. (Indian)      & 72.12 & Spanish (MX)      & 78.81 \\
French             & 83.39 & Tagalog           & 81.09 \\
French (CA)         & 78.56 & Thai$^\dagger$    & 83.67 \\
German             & 84.23 & Turkish           & 92.97 \\
Italian            & 78.16 & Urdu              & 89.63 \\
Japanese$^\dagger$ & 81.46 & Vietnamese        & 71.81 \\
Korean$^\dagger$   & 81.33 &                   &       \\
\midrule
\multicolumn{4}{@{}l}{Baseline average (21 categories): \textbf{79.15}} \\
\multicolumn{4}{@{}l}{\emph{Our system} (macro, 21 categories): \textbf{29.27}} \\
\bottomrule
\end{tabular}
\par\vspace{2pt}
{\footnotesize $^\dagger$ scored with tcpCER.}
\end{table}

\subsection{Cascaded-system ablations}
\label{ssec:ablation}
Table~\ref{tab:ablation} summarizes the main component ablations on the
full Development set (\(150\) conversations, macro tcpMER over the official
\(21\) language/accent categories). \textbf{All four experiments are
scored under the same official protocol} (NFKC + strict symmetric
normalization, collar \(5\,\)s). Following the incremental ablation style of
\cite{ding2022personalvad}, each experiment adds or changes one pipeline
stage while keeping the same LoRA-adapted ASR checkpoint and scoring
procedure.

Exp1 is an end-to-end-style baseline: FunASR FSMN-VAD cuts the audio,
the recognizer emits \texttt{<sc>} turn markers, and speakers are assigned
by alternating \texttt{O1}/\texttt{O2} at each marker---without neural
segmentation or embedding-based clustering. Its macro tcpMER is
\(141.2\%\) (values above \(100\%\) are possible when insertions dominate
under the time-constrained metric). Exp2--Exp4 replace this with a cascaded design: a neural
segmentation model (\texttt{pyannote/segmentation-3.0} or DiariZen-Large-s80)
followed by CAM++ embedding
clustering with two speakers forced. This change alone lowers tcpMER from
\(141.2\%\) to \(35.3\%\) (Exp2), showing that speaker identity cannot be
inferred reliably from turn markers alone. Within the cascaded family,
disabling overlap-aware segmentation (Exp3, \(30.6\%\)) improves over
Exp2 (\(35.3\%\)) because overlapped regions are otherwise transcribed
twice; switching the segmenter to DiariZen-Large-s80 (Exp4, \textbf{29.27\%})
yields a further \(1.3\) point gain. \textbf{Exp4 is the configuration
submitted to the challenge.}

\begin{table}[t]
\centering
\caption{Incremental ablations on the Development set (macro tcpMER over
\(21\) categories; official NFKC + strict normalization). Lower is
better.}
\label{tab:ablation}
\small
\begin{tabular}{@{}lll@{}}
\toprule
Exp & System & tcpMER (\%) \\
\midrule
Exp1 & End-to-end: VAD + ASR & 141.2 \\
Exp2 & + \texttt{pyannote/seg.-3.0}, CAM++ & 35.3 \\
Exp3 & + overlap disabled & 30.6 \\
Exp4 & + DiariZen-Large-s80 (repl.\ pyannote) & \textbf{29.27} \\
\bottomrule
\end{tabular}
\end{table}

\subsection{Leaderboard submission and error analysis}
Under team name \texttt{fangshuming}, our Evaluation-set submission scored
a tcpMER of \textbf{50.23\%}, markedly higher than the \(29.27\%\) macro tcpMER on
the Development set. We ruled out several formatting artifacts: speaker-ID
style (\texttt{1/2} vs \texttt{O1/O2}) is score-neutral under tcpMER, and
overlap is negligible on Evaluation (only two overlapping segment pairs in
\(20{,}914\) segments). Although Evaluation references are not released
and we cannot analyze per-utterance errors, the observed behavior
suggests that the remaining gap is likely due to domain shift: the
Development set was used to tune pipeline hyperparameters, while
Evaluation differs in
speaker, channel, and language mix---especially on already difficult
cohorts (e.g.\ Turkish, French (CA)). We also tried an automatic
translation / ``de-code-switching'' post-step, which \emph{degraded}
tcpMER because references retain code-switched words; it was not used in
the final submission.

\section{Discussion}
Our Development-set ablations suggest three practical lessons for building
systems under the challenge tcpMER metric:

\noindent\textbf{Overlap-aware segmentation.} Overlap detection is
commonly used to improve diarization recall but hurt our tcpMER score:
duplicated overlap regions are
transcribed twice and count as errors under a time-constrained metric.
We therefore disable overlap in the submitted system (Exp3/Exp4).

\noindent\textbf{Speaker embedding.} Segmentation alone does not identify
\emph{who} is speaking. On the Development set, \texttt{<sc>}-only
assignment (Exp1, \(141.2\%\)) remains far above cascaded CAM++
clustering (Exp3, \(30.6\%\)); we keep a separate embedding stage in the
submitted system.

\noindent\textbf{Metric optimization.} Choices that improve diarization
recall (e.g.\ overlap) and choices that improve the challenge ranking
metric are not
the same problem; component choices should be validated on tcpMER
directly.

\section{Conclusion}
We presented our Task~1 submission to the 2nd MLC-SLM Challenge: a
cascaded system with DiariZen-Large-s80 segmentation, CAM++ two-speaker clustering,
and LoRA-adapted \texttt{omniASR\_LLM\_7B\_v2}. On the official Development set it
reaches \(29.27\%\) macro tcpMER (\(50.23\%\) on Evaluation). Development
ablations showed that (i)~a separate embedding-clustering stage is
essential compared with \texttt{<sc>}-only speaker assignment, and
(ii)~overlap-aware segmentation should be turned off for this metric.
Exp4 in Table~\ref{tab:ablation} is the submitted configuration.
These observations may help guide future engineering choices for
multilingual conversational speech systems evaluated under tcpMER.

\bibliographystyle{IEEEtran}
\bibliography{refs}

\end{document}